\documentclass[journal]{IEEEtran}
\usepackage{etoolbox}
\makeatletter
\patchcmd{\@makecaption}
  {\scshape}
  {}
  {}
  {}
\makeatother

\usepackage{caption}
\usepackage{url}
\usepackage{hyperref}
\usepackage[super]{nth}
\usepackage{xspace}
\usepackage{verbatim}
\usepackage{multirow}
\usepackage{booktabs}
\usepackage{enumitem} 
\usepackage{subcaption}
\usepackage{cite}
\usepackage[table,xcdraw]{xcolor}
\usepackage{array}
\usepackage{tabularx}
\newcolumntype{Y}{>{\centering\arraybackslash}X}

\usepackage{graphicx}
\usepackage{siunitx}
\usepackage{amssymb,amsmath,bm}
\usepackage{textcomp}
\usepackage{color}
\usepackage{ulem} 
\ifCLASSINFOpdf
\else
\fi

\usepackage{color}
\usepackage{ulem}

\newcommand{\edit}[1]{\textcolor{black}{#1}}

\begin{document}
%
\title{Wavehax: Aliasing-Free Neural Waveform Synthesis \\ Based on 2D Convolution and Harmonic Prior \\ for Reliable Complex Spectrogram Estimation}
%
%
%

\author{Reo Yoneyama,
        Atsushi Miyashita,
        Ryuichi Yamamoto,
        and~Tomoki Toda,~\IEEEmembership{Member,~IEEE}
\thanks{Manuscript received xxx xx, 2020; revised xxx xx, 2020. This work was supported in part by the Japan Science and Technology Agency (JST), Precursory Research for Embryonic Science and Technology (PRESTO) under Grant JPMJPR1657, in part by the JST, CREST under Grant JPMJCR19A3, and in part by the Japan Society for the Promotion of Science (JSPS) Grants-in-Aid for Scientific Research (KAKENHI) under Grant 24KJ1236.}
\thanks{R. Yoneyama is with Graduate School of Informatics, Nagoya University, Japan (e-mail: yoneyama.reo@g.sp.m.is.nagoya-u.ac.jp).}
\thanks{A. Miyashita is with Graduate School of Informatics, Nagoya University, Japan (e-mail: miyashita.atsushi@g.sp.m.is.nagoya-u.ac.jp).}
\thanks{R. Yamamoto is with LY Corporation, Japan, and Graduate School of Informatics, Nagoya University, Japan (e-mail: ryuichi.yamamoto@lycorp.co.jp).}
\thanks{T. Toda is with Information Technology Center, Nagoya University, Japan (e-mail: tomoki@icts.nagoya-u.ac.jp).}
}

%
%

\markboth{Journal of \LaTeX\ Class Files,~Vol.~14, No.~8, August~2015}%
{Shell \MakeLowercase{\textit{et al.}}: Bare Demo of IEEEtran.cls for IEEE Journals}
%



\maketitle

\begin{abstract}
Neural vocoders often struggle with aliasing in latent feature spaces, caused by time-domain nonlinear operations and resampling layers.
Aliasing folds high-frequency components into the low-frequency range, making aliased and original frequency components indistinguishable and introducing two practical issues.
First, aliasing complicates the waveform generation process, as the subsequent layers must address these aliasing effects, increasing the computational complexity.
Second, it limits extrapolation performance, particularly in handling high fundamental frequencies, which degrades the perceptual quality of generated speech waveforms.
This paper demonstrates that 1) time-domain nonlinear operations inevitably introduce aliasing but provide a strong inductive bias for harmonic generation, and 2) time-frequency-domain processing can achieve aliasing-free waveform synthesis but lacks the inductive bias for effective harmonic generation.
Building on this insight, we propose Wavehax, an aliasing-free neural WAVEform generator that integrates 2D convolution and \edit{a HArmonic prior} for reliable Complex Spectrogram estimation.
Experimental results show that Wavehax achieves speech quality comparable to existing high-fidelity neural vocoders and exhibits exceptional robustness in scenarios requiring high fundamental frequency extrapolation, where aliasing effects become typically severe.
Moreover, Wavehax requires less than 5\% of the multiply-accumulate operations and model parameters compared to HiFi-GAN V1, while achieving over four times faster CPU inference speed.
\end{abstract}


%
\IEEEpeerreviewmaketitle

\section{Introduction}

\IEEEPARstart{V}OCODERS \cite{vocoder-1939} are models that synthesize audio waveforms from input acoustic features.
Specifically, vocoders utilizing deep neural networks (DNNs) are called neural vocoders.
With advances in deep learning, neural vocoders have achieved high-fidelity speech synthesis often indistinguishable from natural human speech.
Historically, the development of neural vocoders has focused primarily on time-domain models, including autoregressive vocoders \cite{wavenet, wavenet-vocoder, wavernn, lpcnet, fb-lpcnet}, normalizing flow \cite{flow} vocoders \cite{parallel-wavenet, clarinet, pwngan, waveflow, waveglow}, generative adversarial networks (GAN) \cite{gan} vocoders \cite{pwg, melgan, mb-melgan, hifigan, ms-hifigan, bigvgan}, and denoising diffusion probabilistic model \cite{ddpm} vocoders \cite{wavegrad, diffwave, priorgrad, specgrad}.
Among these, GAN-based methods have been extensively studied for their fast synthesis speed, compact generator, and high-fidelity outputs.
These methods typically employ one-dimensional (1D) convolutional neural networks (CNNs) followed by nonlinear activation functions, such as rectified linear units (ReLU) \cite{relu}.
Additionally, some methods \cite{melgan, mb-melgan, hifigan, ms-hifigan, bigvgan} utilize multiple upsampling layers to efficiently convert low-resolution acoustic features into high-resolution waveforms.

Despite advancements in neural vocoders, their architectural design has received limited attention from a signal-processing perspective, particularly regarding aliasing in latent feature spaces caused by nonlinear operations \cite{stylegan3, bigvgan} and resampling layers \cite{upsample-artifact}.
Aliasing folds high-frequency components back into the low-frequency range, making the aliased and original low-frequency components indistinguishable, and introduces undesirable distortion, known as aliasing artifacts.
These phenomena induce two practical problems.
First, aliasing complicates the synthesis process because the subsequent layers must address the aliasing effects, resulting in increased computational complexity.
Second, aliasing limits extrapolation performances, particularly for fundamental frequency ($\text{F}_0$) values, degrading the perceptual quality of generated waveforms \cite{qpnet-journal, qppwg-journal, pap-gan, eval_extrapolation}.
This problem is especially relevant in applications such as text-to-speech, voice conversion for new speakers, and singing voice synthesis involving pitch shifts.

Lee et al. \cite{bigvgan} highlighted that while nonlinear operations create specific frequency components, their application to discrete-time signals inevitably results in aliasing due to the frequency band limitations imposed by the Nyquist–Shannon sampling theorem \cite{sampling_theorem1, sampling_theorem2}.
To mitigate the aliasing caused by nonlinear operations, they employed temporal upsampling to extend the representable frequency band before applying nonlinear operations.
Pons et al. \cite{upsample-artifact} investigated upsampling artifacts, including aliasing, across various upsampling techniques in neural vocoders.
They also suggested that network training plays a crucial role in compensating for these artifacts.
While these studies provided valuable insights into mitigating aliasing in neural vocoders, completely eliminating aliasing effects remains an open challenge.
Additionally, countermeasures that rely on model training do not guarantee robust anti-aliasing performance when applied to unseen data.

This paper demonstrates that 1) time-domain nonlinear operations inevitably introduce aliasing but provide a strong inductive bias for harmonic generation (where harmonics are integer multiples of a fundamental frequency), and 2) time-frequency-domain processing can achieve aliasing-free waveform synthesis but lacks the inductive bias for effective harmonic generation.
Based on this insight, we propose Wavehax, an aliasing-free neural vocoder that estimates complex spectrograms and converts them into time-domain waveforms via the short-time Fourier transform (STFT).
\edit{Wavehax employs a novel integration of 2D CNNs with a complex spectrogram derived from a harmonic signal, which is crucial for high-fidelity and robust complex spectrogram estimation.}

Importantly, this study distinguishes itself from previous works on neural vocoders estimating complex spectrograms \cite{istftnet, istftnet2, hiftnet, apnet, lightvoc, vocos}, which primarily emphasize the computational efficiency of time-frequency-domain processing.
In contrast, our approach prioritizes aliasing-free waveform synthesis, demonstrating its theoretical and practical benefits.
Experimental results show that Wavehax achieves speech quality comparable to existing high-fidelity neural vocoders and exhibits exceptional robustness in high $\text{F}_0$ extrapolation scenarios, where aliasing effects become typically severe.
Moreover, Wavehax requires less than 5\% of the multiply-accumulate operations and model parameters compared to HiFiGAN V1 \cite{hifigan}, while achieving over four times faster CPU inference speed.
Audio samples and code are available from our demo site\footnote{\url{https://chomeyama.github.io/wavehax-demo/}}.

\section{Related Work}
\label{sec: related work}

In this section, we introduce time-domain neural vocoders and causes of aliasing.
We then outline anti-aliasing techniques employed in neural vocoders.
Finally, we review current neural vocoders that utilize complex spectrogram estimation and highlight the drawbacks that our study addresses.

\subsection{Time-domain neural vocoders}
\label{sec2: vocoders}

Time domain vocoders such as WaveNet vocoder \cite{wavenet-vocoder} and Parallel WaveGAN (PWG) \cite{pwg} generate speech waveforms by iteratively applying 1D dilated CNNs and nonlinear activation functions in the waveform domain.
These vocoders operate at a fixed temporal resolution from input to output.
However, audio signals have high-temporal resolutions (e.g., 24 K samples per second), which increases computational complexity.
Therefore, this type of neural vocoder with a considerable model size has difficulty achieving sufficient synthesis speed, which is crucial for real-time applications.
Other neural vocoders, such as MelGAN \cite{melgan} and HiFi-GAN \cite{hifigan}, progressively upsample low-temporal-resolution acoustic features to convert them into high-temporal-resolution waveforms.
This strategy facilitates efficient waveform synthesis thanks to the low computational cost of low-temporal-resolution processing, with a significant advantage in synthesis speed.
Additionally, upsampling-based vocoders can easily expand the receptive field.

\subsection{Aliasing due to nonlinear operation}
\label{ssec: aliasing due to nonlinear operation}

\cite{stylegan3} and \cite{bigvgan} have demonstrated that pointwise nonlinear operations are essential for generating new frequency components in neural networks.
A simple example is the ReLU \cite{relu} activation function, which operates as the multiplication of dynamically determined rectangular windows based on the input signal.
When applied to a sinusoidal signal with angular frequency $\omega$ defined for continuous time $t \in \mathbb{R}$, the ReLU function generates an infinite number of harmonics, as shown in its Fourier expansion:
\begin{equation} \label{eq: rectified sine}
  \mathrm{relu}(\mathrm{sin}(\omega t)) = \frac{1}{\pi} + \frac{\mathrm{sin}(\omega t)}{2} - \sum_{k=1}^{\infty} \frac{2 \, \mathrm{cos}(2k\omega t)}{\pi(2k - 1)(2k + 1)}.
\end{equation}
This equation provides a frequency domain interpretation in which the input frequency characteristics are convolved with $\mathrm{sinc}$ functions that possess infinite-length side lobes.
As this example illustrates, nonlinear operations are critical for neural vocoders to represent complex functions while also providing a strong inductive bias for harmonic generation (as discussed in Section~\ref{sssec: prior for time-domain models}).
However, applying such pointwise operations to discrete-time signals can induce aliasing due to the frequency band limitations imposed by the Nyquist–Shannon sampling theorem \cite{sampling_theorem1, sampling_theorem2}.

To address this issue, BigVGAN \cite{bigvgan} performs nonlinear operations in an anti-aliased manner, following the approach in StyleGAN3 \cite{stylegan3}.
\cite{stylegan3} highlights that aliasing within neural networks can blur high- and low-frequency components, complicating the control of coarse and fine features separately.
Their approach involves temporally upsampling input signals before applying nonlinear operations to extend the representable frequency range dictated by the sampling theorem.
This technique enables \cite{bigvgan, stylegan3} to effectively reduce aliasing and enhance the quality of the generated outputs.

\subsection{Aliasing due to upsampling layers}
\label{sec2: aliasing due to upsampling layers}

Although upsampling layers in neural vocoders facilitate efficient waveform generation, they often introduce upsampling artifacts, including aliasing.
Pons et al. \cite{upsample-artifact} comprehensively analyzed these artifacts across various upsampling techniques, such as transposed CNNs, subpixel CNNs \cite{subpixel-cnn}, and interpolation-based upsamplers.
They identified tonal artifacts, characterized by repetitive patterns, as a major factor degrading perceptual quality, especially induced by transposed and subpixel CNNs.
While interpolation-based upsamplers can avoid tonal artifacts, they tend to introduce filtering artifacts due to specific frequency responses of filters (e.g., the triangular filter for linear interpolation).
Moreover, upsampling discrete signals introduces aliasing (i.e., spectral replicas), which can propagate artifacts from previous layers.
Pons et al. \cite{upsample-artifact} also suggested that network training can mitigate these artifacts by adjusting neural network parameters.

Shang et al. \cite{ssim-alias} addressed aliasing artifacts from the perspective of artifact detection.
They focused on the spectral symmetry caused by aliasing and proposed a learning criterion based on the structural similarity index \cite{ssim}, a common metric for image quality evaluation.
However, mitigating aliasing artifacts that depend on model training does not guarantee robust performance on unseen data, particularly when encountering unseen $\text{F}_0$ values.
Consequently, we argue that anti-aliasing strategies independent of training or specific criteria are preferable for achieving robust performance.

\subsection{Time-frequency domain neural vocoders}

Several studies have developed neural vocoders that estimate time-frequency representations rather than directly generating audio waveforms.
For example, SpecGAN \cite{specgan} estimates log amplitude spectrograms.
Similarly, GANstrument \cite{ganstrument} estimates log amplitude spectrograms with a mel-frequency scale.
Estimating phase spectrograms is particularly challenging due to their randomness and the ambiguity caused by multiples of $2\pi$.
As a result, these methods rely on the Griffin-Lim algorithm (GLA) \cite{gla} to convert spectrograms into audio waveforms.
However, using GLA can lead to unnatural audio, as it relies solely on waveform-spectrogram consistency to estimate the phase spectrograms.

\edit{HiNet \cite{hinet} hierarchically estimates the log amplitude and phase spectrogram, ultimately reconstructing the waveform using inverse STFT (iSTFT).
It estimates a log amplitude spectrogram using a feed-forward network and conditions a neural source-filter (NSF) \cite{nsf-journal} model with the log amplitude spectrogram to generate a time-domain waveform, from which the phase spectrogram is extracted via STFT.
The excitation generation module of the NSF constructs harmonic signals by accumulating phase rotation angles for each time step, determined by $\text{F}_0$.
This process ensures a coherent phase structure of the output waveforms, aiding efficient phase spectrogram estimation and allowing HiNet to sidestep the challenges of direct phase estimation.
However, the phase spectrogram derived from time-domain processing involving nonlinear operations can introduce aliasing.}

ISTFTNet \cite{istftnet, istftnet2} is the first vocoder to directly estimate complex spectrograms.
It converts low-temporal-resolution mel-spectrograms into high-temporal-resolution, low-frequency-resolution complex spectrograms using multiple transposed CNNs and HiFi-GAN-based residual blocks.
The final iSTFT transforms these complex spectrograms into speech waveforms.
Although ISTFTNet facilitates fast waveform generation, its reliance on upsampling layers introduces upsampling artifacts (see Section~\ref{sec2: aliasing due to upsampling layers}).
Additionally, its limited frequency bins in the complex spectrogram lead to insufficient frequency resolution, complicating optimal STFT configurations for different datasets.

HiFTNet \cite{hiftnet} incorporates source-filter modeling \cite{source-filter} into iSTFTNet, extending it with an $\text{F}_0$ estimation network and a periodic signal generation scheme based on harmonic-plus-noise NSF \cite{nsf-journal}.
The architecture is further enhanced with techniques like the Snake activation function \cite{snake} (as used in BigVGAN) and truncated pointwise relativistic loss \cite{truncated_gan}.
However, it faces challenges similar to iSTFTNet, including aliasing artifacts from multiple resampling layers and insufficient frequency resolution due to limited frequency bins.

APNet \cite{apnet} estimates complex spectrograms without upsampling layers.
To address the difficulty of direct phase spectrogram estimation, APNet is trained to align multiple aspects between the target and generated complex spectrograms, including log amplitude spectrogram, real and imaginary parts, instantaneous frequency, group delay, and phase time difference.
However, relying on explicit distance losses that assume a one-to-one relationship between complex spectrograms and waveforms can be suboptimal due to the inherent redundancy in complex spectrograms.

In contrast, LightVoc \cite{lightvoc} and Vocos \cite{vocos} adopt primary training objectives: mel-spectrogram reconstruction loss, adversarial loss, and feature matching loss, as common GAN-based vocoders.
They employ sophisticated architectures such as ConvNeXt \cite{convnext} and Conformer \cite{conformer}.
Especially, Vocos achieves faster waveform generation than iSTFTNet while maintaining comparable audio quality to BigVGAN through training on a large-scale dataset.
However, this paper shows that while Vocos leverages a large number of parameters for high expressiveness, it struggles to robustly handle unseen $\text{F}_0$ values (Section~\ref{sec: exA}).

\section{Theoretical Background}
\label{sec: background}

This section discusses the fundamental properties of time, frequency, and time-frequency domain operations in neural vocoders from a signal-processing viewpoint, particularly focusing on aliasing.

\subsection{Time-domain processing}
\label{ssec: time-domain processing}

We start by discussing time-domain convolution and nonlinear operations, focusing on why nonlinear operations in the time domain, such as the ReLU activation function \cite{relu}, cause aliasing.
We then examine the anti-aliased nonlinear operation \cite{stylegan3, bigvgan} and highlight their limitations.

\subsubsection{Time-domain convolution}
\label{sssec: time-domain convolution}

Let $\bm{x} \in \mathbb{R}^N$ be a discrete-time signal defined at each time step $n = 0, 1, \ldots, N-1$.
We denote the element at time step $n$ as $\bm{x}[n]$.
The frequency representation of $\bm{x}$ is given by $\mathcal{F}\{ \bm{x} \} \in \mathbb{C}^N$, which is obtained via the discrete-time Fourier transform (DFT) $\mathcal{F}$ as follows:
\begin{equation} \label{eq: dft}
    \mathcal{F}\{ \bm{x} \}[k] = \sum_{n=0}^{N-1} \bm{x}[n] \, \mathrm{exp}(-j 2 \pi \frac{k}{N} n),
\end{equation}
where $0 \leq k < N$ represents the frequency indices.
Let $\bm{h} \in \mathbb{R}^M$ be a filter of length $M$.
The time-domain convolution can be expressed as:
\begin{equation} \label{eq: time-domain convolution}
(\bm{x} * \bm{h})[n] = \sum_{m=0}^{M-1} \bm{x}[\, (n + m) ~\mathrm{mod}~ N] \, \bm{h}[m],
\end{equation}
which can be implemented as a 1D convolutional layer in neural networks.
To simplify the equations, we assume periodic padding, although zero padding or reflection padding can also be applied.
According to the convolution theorem, this time-domain convolution modifies each frequency component $\mathcal{F}\{ \bm{x} \}[k]$ depending on $\mathcal{F}\{ \bm{h} \}$ in the frequency domain.
Thus, it operates as a linear system with a frequency response characterized by the magnitude $|\mathcal{F}\{ \tilde{\bm{h}} \}|$ and phase $\angle \mathcal{F}\{ \tilde{\bm{h}} \}$.

\subsubsection{Time-domain nonlinear operation}
\label{sssec: time-domain nonlinear operation}

Let $f: \mathbb{R} \rightarrow \mathbb{R}$ be an arbitrary pointwise nonlinear function.
From the pointwise property, the resulting signal from applying $f$ to each element of the input signal $\bm{x}$ can be expressed as $f(\bm{x})[n] = f(\bm{x}[n])$.
To facilitate the analysis of the anti-aliased nonlinear operation discussed in Section~\ref{sssec: anti-aliased nonlinear operation}, we provide an equivalent interpretation and implementation of the pointwise nonlinear operation $f(\bm{x})$ as a vector multiplication.
Specifically, we introduce a coefficient signal $\bm{a}$, so that the function $f(\bm{x})$ can be expressed as the Hadamard product of $\bm{x}$ and $\bm{a}$, i.e., $f(\bm{x}) = \bm{x} \odot \bm{a}$, where $\bm{a}$ is defined as:
\begin{equation} \label{eq: time-domain nonlinear}
    \bm{a}[n] =
    \begin{cases}
        f(\bm{x}[n]) ~ / ~ \bm{x}[n] & \mbox{if} ~ \bm{x}[n] \neq 0 \\
        0 & \mbox{otherwise}.
    \end{cases}
\end{equation}
According to the inverse convolution theorem, multiplying two discrete-time signals in the time domain corresponds to the circular convolution of their spectra in the frequency domain.
Therefore, the spectrum of $\bm{x} \odot \bm{a}$ is given by:
\begin{equation} \label{eq: circular conv}
\begin{aligned}
    \mathcal{F}\{ \bm{x} \odot \bm{a} & \}[k] = (\mathcal{F} \{ \bm{x} \} * \mathcal{F} \{ \bm{a} \}) [k] \\
    &= \sum_{m=0}^{N-1} \mathcal{F} \{ \bm{x} \}[\,(k + m) ~\mathrm{mod}~ N] ~ \mathcal{F} \{ \bm{a} \}[m].
    \end{aligned}
\end{equation}
This equation indicates that high-frequency components with indices exceeding $N$ fold back into the lower-frequency band due to the periodic nature of the discrete Fourier transform, resulting in overlapping spectra.
If the sampling frequency of $\bm{x}$ is insufficient to represent the bandwidth required for $f(\bm{x})$, aliasing occurs.

\subsection{Anti-aliased nonlinear operation}
\label{sssec: anti-aliased nonlinear operation}

\begin{figure}[t]
    \begin{center}
    \includegraphics[width=0.96\columnwidth]{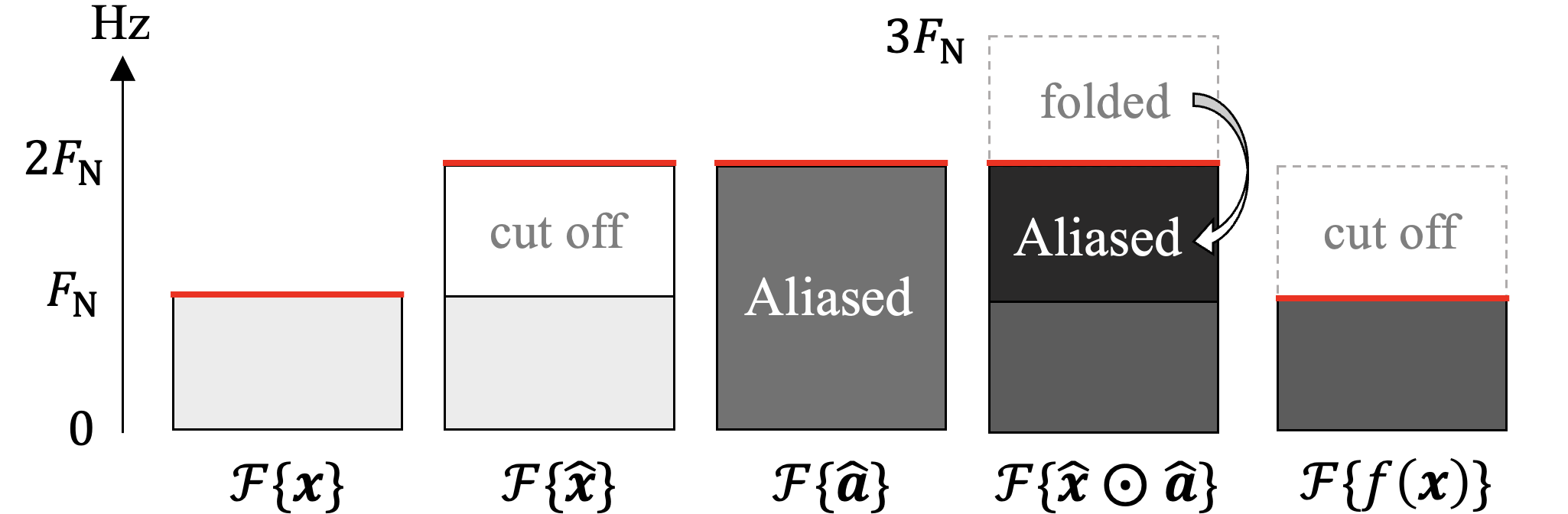}
    \end{center}
    \vspace{-1mm}
    \caption{\small Frequency domain representations of each signal in the anti-aliased nonlinear operation \cite{bigvgan} are depicted. The output signal $f(\bm{x})$, shown on the right, is obtained through the process described in Eq.~(\ref{eq: anf2}). The red lines indicate the Nyquist frequencies.}
    \label{fig: anti-alias}
    \vspace{-1mm}
\end{figure}

BigVGAN \cite{bigvgan} incorporates the anti-aliased nonlinear operation \cite{stylegan3} to mitigate aliasing effects.
After explaining the original formulation of this operation, we present an equivalent interpretation and implementation based on vector multiplication, as described in Section~\ref{sssec: time-domain nonlinear operation}.
Finally, we discuss the limitations and drawbacks of the anti-aliased nonlinear operation.

\subsubsection{Original formulation}
\label{sssec: original formulation}

The anti-aliased nonlinear operation first upsamples the input signal $\bm{x}$ by a factor of two, applies the nonlinear operation $f$, and then downsamples the signal back to its original temporal resolution after low-pass filtering.
Defining $F_\text{N}$ as the Nyquist frequency associated with $\bm{x}$, the resulting signal $f(\bm{x})$ is obtained as follows:
\begin{equation} \label{eq: anf1}
    \hat{\bm{x}} = \mathrm{lowpass}(\, \mathrm{resample}(\bm{x}, 2), F_\text{N}),
\end{equation}
\begin{equation} \label{eq: anf2}
    f(\bm{x}) = \mathrm{resample}(\, \mathrm{lowpass}(f(\hat{\bm{x}}), F_\text{N}), 0.5),
\end{equation}
where $\mathrm{lowpass}(\bm{v}, c)$ represents low-pass filtering that retains only frequencies below $c$, and $\mathrm{resample}(\bm{v}, c)$ denotes resampling the input signal $\bm{v}$ by a factor of $c$.
Note that $f$ is applied to the upsampled signal $\hat{\bm{x}} \in \mathbb{R}^{2N}$, which has a Nyquist frequency of $2F_\text{N}$.

\subsubsection{Equivalent interpretation and implementation}
\label{sssec: equivalent interpretation and implementation}

Building on the discussion in Section~\ref{sssec: time-domain nonlinear operation}, we can derive an alternative implementation for computing $f(\hat{\bm{x}})$ in Eq.~(\ref{eq: anf2}) using a coefficient signal $\hat{\bm{a}} \in \mathbb{R}^{2N}$ as follows:
\begin{equation} \label{eq: a_hat}
    \hat{\bm{a}}[n] =
    \begin{cases}
        f(\hat{\bm{x}}[n]) ~ / ~ \hat{\bm{x}}[n] & \mbox{if} ~ \hat{\bm{x}}[n] \neq 0 \\
        0 & \mbox{otherwise},
    \end{cases}
\end{equation}
\begin{equation} \label{eq: reformed ano}
    f(\hat{\bm{x}}) = \hat{\bm{x}} \odot \hat{\bm{a}}.
\end{equation}
This reformulation clarifies the advantages of the anti-aliased nonlinear operation.
Similar to Eq.~(\ref{eq: circular conv}), the spectrum of $\hat{\bm{x}} \odot \hat{\bm{a}}$ is given by the circular convolution of the spectra of $\hat{\bm{x}}$ and $\hat{\bm{a}}$:
\begin{equation} \label{eq: anf circular conv}
\begin{aligned}
    \mathcal{F} \{ \hat{\bm{x}} \odot \hat{\bm{a}} & \} = (\mathcal{F} \{ \hat{\bm{x}} \} * \mathcal{F} \{ \hat{\bm{a}} \}) [k] \\
    &= \sum_{m=0}^{2N-1} \mathcal{F} \{ \hat{\bm{x}} \}[\, (k + m) ~\mathrm{mod}~ 2N] ~ \mathcal{F} \{ \hat{\bm{a}} \}[m].
\end{aligned}
\end{equation}
In this context, the bandwidth of $\hat{\bm{x}}$ is limited to $2F_\text{N}$, which results in spectral overlap in the frequency range from $F_\text{N}$ to $2F_\text{N}$, as shown in Fig.~\ref{fig: anti-alias}.
As indicated in Eq.~(\ref{eq: anf2}), a low-pass filter removes this overlapping frequency range, effectively mitigating aliasing in the resulting signal $f(\bm{x})$.

\begin{figure}[t]
    \begin{center}
    \includegraphics[width=0.98\columnwidth]{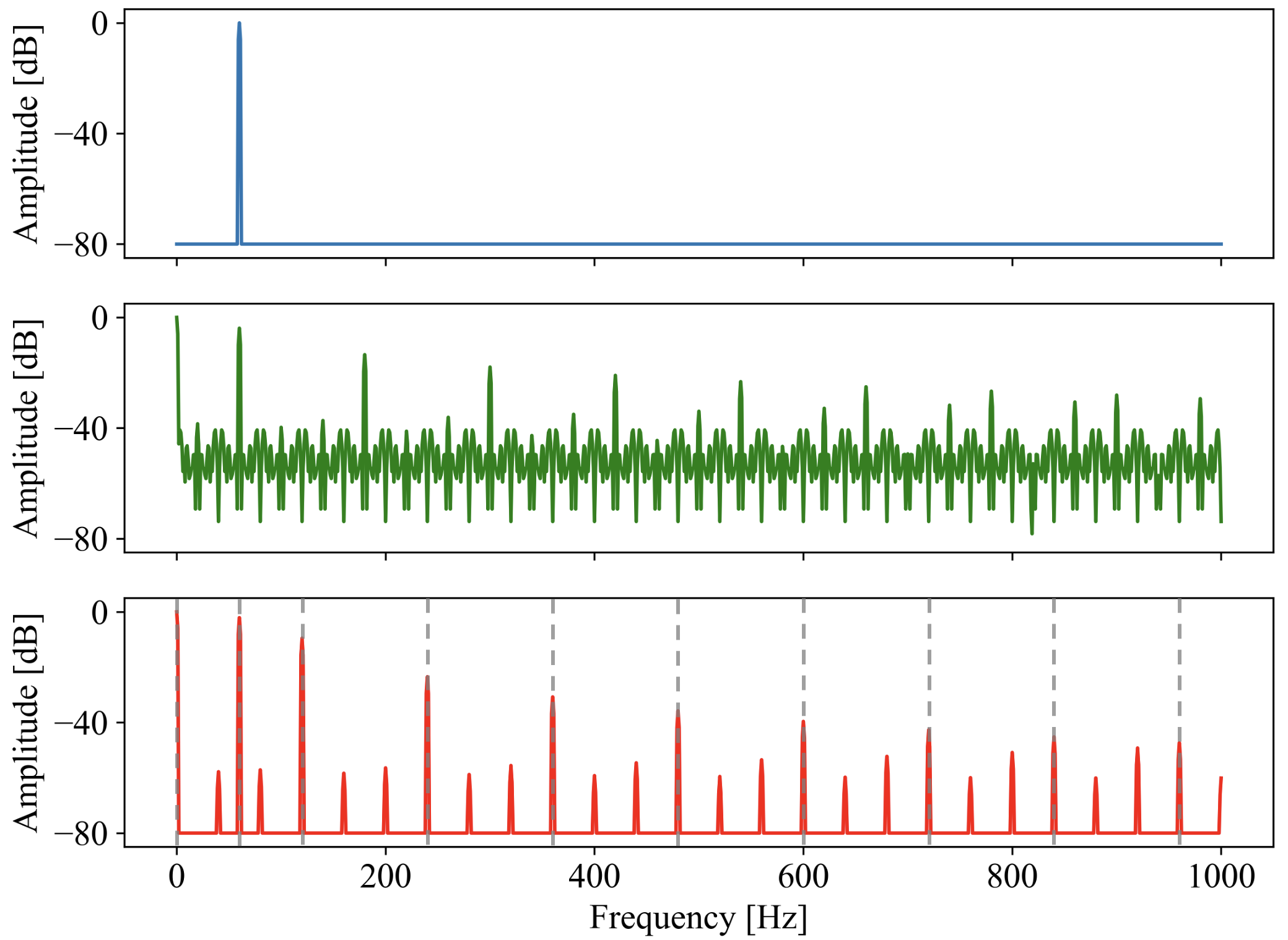}
    \end{center}
    \vspace{-1mm}
    \caption{\small Amplitude spectra in dB of $\hat{\bm{x}}$ (blue), $\hat{\bm{a}}$ (green), and $\hat{\bm{x}} \odot \hat{\bm{a}}$ (red) are shown for a 60 Hz sinusoidal signal $\bm{x}$ when the ReLU \cite{relu} function is applied. The sampling frequency of the original signal $\bm{x}$ is 1 kHz. The dotted lines mark expected harmonic frequencies from Eq.~(\ref{eq: rectified sine}). Despite the use of an anti-aliasing nonlinear operation~\cite{bigvgan}, aliasing artifacts are still clearly observed in the green and red spectra.}
    \label{fig: spectra}
    \vspace{-1mm}
\end{figure}

\subsubsection{Limitations and drawbacks}
\label{sssec: problems of anti-aliased nonlinear operation}

While the anti-aliased nonlinear operation effectively reduces aliasing, it has notable limitations.
First, it does not prevent aliasing introduced into the coefficient signal $\hat{\bm{a}}$ in Eq.~(\ref{eq: a_hat}).
For instance, consider applying the ReLU activation function using the anti-aliased nonlinear operation.
The rectified version of the input signal $\hat{\bm{x}}$ can be expressed as the product of $\hat{\bm{x}}$ and a coefficient signal $\hat{\bm{a}}$, defined as follows:
\begin{equation} \label{eq: relu}
    \hat{\bm{a}}[n] =
    \begin{cases}
        1 & \mbox{if} ~ \hat{\bm{x}}[n] > 0 \\
        0 & \mbox{otherwise}.
    \end{cases}
\end{equation}
As described in Section \ref{ssec: aliasing due to nonlinear operation}, the ReLU activation function corresponds to multiplying rectangular windows with infinite support in the frequency domain.
However, pointwise computation of $\hat{\bm{a}}$ corresponds to sampling from a continuous-time signal, potentially introducing aliasing into $\hat{\bm{a}}$.
Figure~\ref{fig: spectra} shows that aliasing occurs in both $\hat{\bm{a}}$ and $\hat{\bm{x}} \odot \hat{\bm{a}}$ when ReLU is applied to a sinusoidal signal.
Additionally, as indicated by Eq.~(\ref{eq: rectified sine}), higher $\text{F}_0$ values result in stronger harmonic amplitudes, which makes aliasing more pronounced at higher $\text{F}_0$ and can degrade the quality of synthesized high-pitched audio.
Second, the need for resampling after each nonlinear operation increases computational cost, reducing synthesis speed and limiting real-time applicability.
The effectiveness of the anti-aliased nonlinear operation depends on the frequency cut-off characteristics of low-pass filtering, where better cut-off characteristics require a longer filter length, thereby increasing computational complexity.
Consequently, there is a trade-off between quality and computational efficiency.

\subsection{Frequency-domain processing}
\label{ssec: frequency-domain processing}

We now turn to frequency-domain convolution and nonlinear operations, highlighting how they can effectively avoid aliasing.
In this context, frequency-domain processing refers to operations performed on the spectrum $\mathcal{F} \{ \bm{x} \} \in \mathbb{C}^N$, obtained by applying DFT to the time-domain signal $\bm{x} \in \mathbb{R}^N$.

\subsubsection{Frequency-domain convolution}
\label{sssec: frequency-domain convolution}

Let $\bm{l} \in \mathbb{C}^L$ be a convolution kernel of length $L \leq N$.
If periodic padding is used, the convolution in the frequency domain is given by:
\begin{equation}
(\mathcal{F} \{ \bm{x} \} * \bm{l})[k] = \sum_{m=0}^{L-1} \mathcal{F} \{ \bm{x} \}[\, (k + m) ~\mathrm{mod}~ N] ~ \bm{l}[m].
\end{equation}
Unlike time-domain nonlinear operations, which result in circular convolution in the frequency domain, frequency-domain convolution is a localized process limited by the receptive field size $L$.
If $L$ is sufficiently smaller than the spectrum length $N$, this operation avoids problematic spectral overlaps, thereby effectively avoiding the aliasing.

\subsubsection{Frequency-domain nonlinear operation}
\label{sssec: frequency-domain nonlinear operation}

Consider an arbitrary pointwise nonlinear operation in the frequency domain, denoted as $g: \mathbb{C} \rightarrow \mathbb{C}$.
As discussed in Section \ref{sssec: time-domain nonlinear operation}, the resulting spectrum $g(\mathcal{F} \{ \bm{x} \})[k] = g(\mathcal{F} \{ \bm{x} \} [k])$ can be expressed as the Hadamard product of $\mathcal{F} \{ \bm{x} \}$ and a coefficient vector $\bm{b} \in \mathbb{C}^N$.
That is, $g(\mathcal{F} \{ \bm{x} \}) = \mathcal{F} \{ \bm{x} \} \odot \bm{b}$, where $\bm{b}$ is defined as:
\begin{equation} \label{eq: nonlinear}
    \bm{b}[k] =
    \begin{cases}
        g(\mathcal{F} \{ \bm{x} \} [k]) ~ / ~ \mathcal{F} \{ \bm{x} \} [k] & \mbox{if} ~ \mathcal{F} \{ \bm{x} \} [k] \neq 0 \\
        0 & \mbox{otherwise}.
    \end{cases}
\end{equation}
This operation independently modifies each element of $\mathcal{F} \{ \bm{x} \}$, similar to time-domain convolution (Section~\ref{sssec: time-domain convolution}).
Although this corresponds to circular convolution in the time domain, potentially causing time-domain overlaps, this paper focuses on spectral overlaps in $\mathcal{F} \{ \bm{x} \}$, recognizing them as a more critical factor contributing to practical drawbacks.
Consequently, frequency-domain nonlinear operations do not introduce problematic spectral overlap in $\bm{x}$.

\subsection{Time-frequency-domain processing}
\label{ssec: time-frequency-domain processing}

Typical audio signals, such as speech or music, are non-stationary and are best represented using time-frequency representations like spectrograms.
To extend our discussion of 1D frequency-domain processing in Section~\ref{ssec: frequency-domain processing} to the 2D time-frequency domain, we consider the spectrogram $\bm{X} \in \mathbb{C}^{M \times K}$, which is obtained by analyzing the input time signal $\bm{x} \in \mathbb{R}^{N}$ via STFT.
Here, $M$ is the number of time frames and $K$ is the number of frequency bins, while $\bm{X}[m, k]$ represents the $k$-th frequency component in the $m$-th time frame.
Unlike 1D convolution in the frequency domain, 2D convolution in the time-frequency domain influences each spectrum along the time axis.
Despite this, the process remains localized and does not cause problematic spectral overlaps.
Meanwhile, pointwise nonlinear operations in the time-frequency domain are applied independently to each spectrum $\bm{X}[m]$, coming down to the discussion in Section~\ref{sssec: frequency-domain nonlinear operation}.
Consequently, time-frequency-domain processing offers advantages over time-domain processing, as it can effectively avoid aliasing.

\section{Proposed Method}
\label{sec: proposed method}

\begin{figure*}[tb]
    \begin{center}
    \includegraphics[width=2\columnwidth]{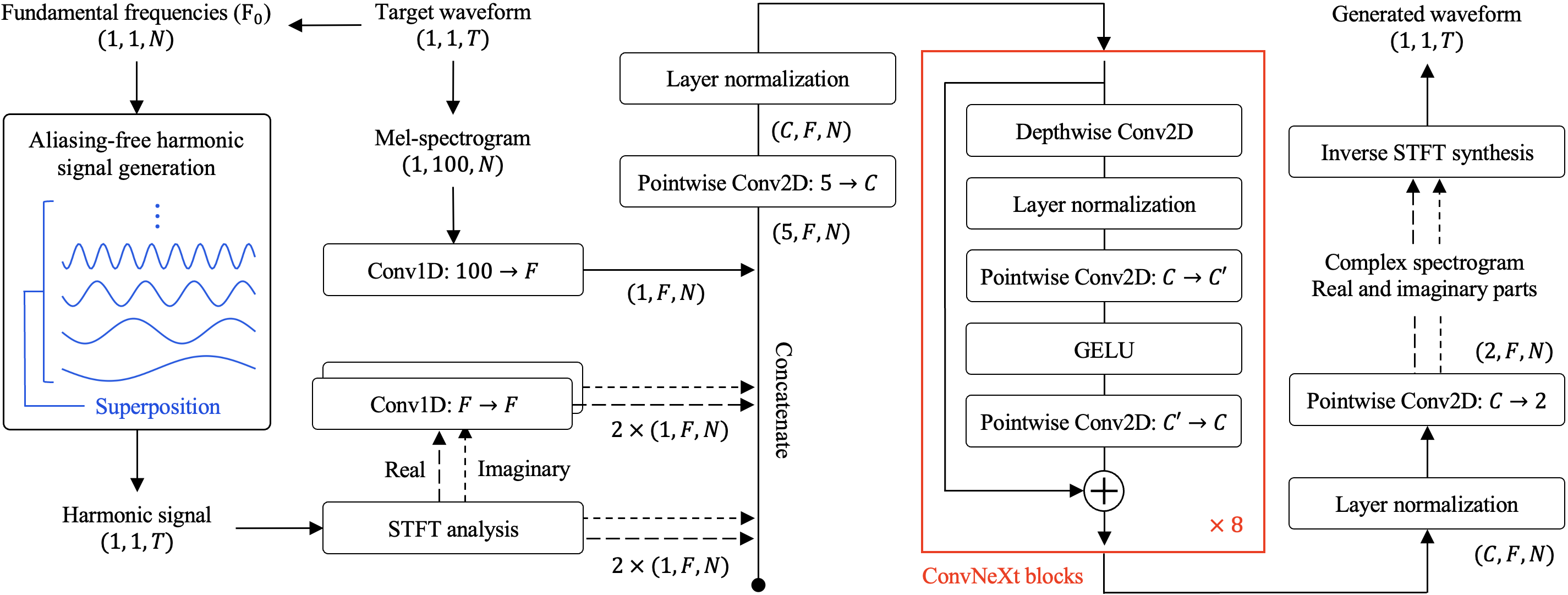}
    \end{center}
    \vspace{-1mm}
    \caption{\small This diagram provides an overview of Wavehax. The kernel width of the 1D convolution is set to 7, while the kernel size of the depthwise convolution is set to 7 $\times$ 7. The numbers of hidden channels, denoted as $C$ and $C'$, are set to 32 and 64, respectively. The number of frequency bins, $F$, is set to 241, calculated as half of the discrete Fourier transform points plus one. $T$ and $N$ represent the number of time steps in the waveforms and time frames in the features, respectively.}
    \label{fig: model}
    \vspace{-1mm}
\end{figure*}

This paper introduces Wavehax, an aliasing-free neural vocoder based on the standard GAN \cite{gan} vocoder framework, designed to avoid aliasing by leveraging time-frequency domain processing, as discussed in Section~\ref{sec: background}.
Additionally, we highlight the importance of integrating 2D CNNs and a harmonic prior for robust, high-fidelity complex spectrogram estimation.
Figure~\ref{fig: model} illustrates an overview of Wavehax.

\subsection{\edit{Harmonic prior}}
\label{ssec: harmonic prior}

\edit{Analogous to prior distributions in Bayesian statistics, we use the term $prior$ to refer to signals considered by the model before observing any training examples.
Intuitively, the prior signal acts as an explicit bias, guiding the model in generating the output waveform\footnote{\edit{Note that GAN-based vocoders, such as PWG \cite{pwg}, can be seen as using a non-informative prior, where Gaussian noise serves as input.
This absence of prior knowledge forces the models to infer the spectral and temporal structures of output waveforms entirely from the training data.}}.
Periodic priors have proven effective for explicit prosodic control in numerous studies, particularly in GAN vocoders \cite{pap-gan, periodnet, nhv, hn-pwg, usfgan-journal, sifigan, firnet}, as well as in other frameworks \cite{nsf-journal, ddsp, periodgrad}.
We further analyze and provide a theoretical explanation of how periodic priors enhance neural vocoders in both the time and time-frequency domains.}

\subsubsection{Periodic prior for time domain models}
\label{sssec: prior for time-domain models}

\begin{figure}[t]
    \begin{center}
    \vspace{-1mm}
    \includegraphics[width=\columnwidth]{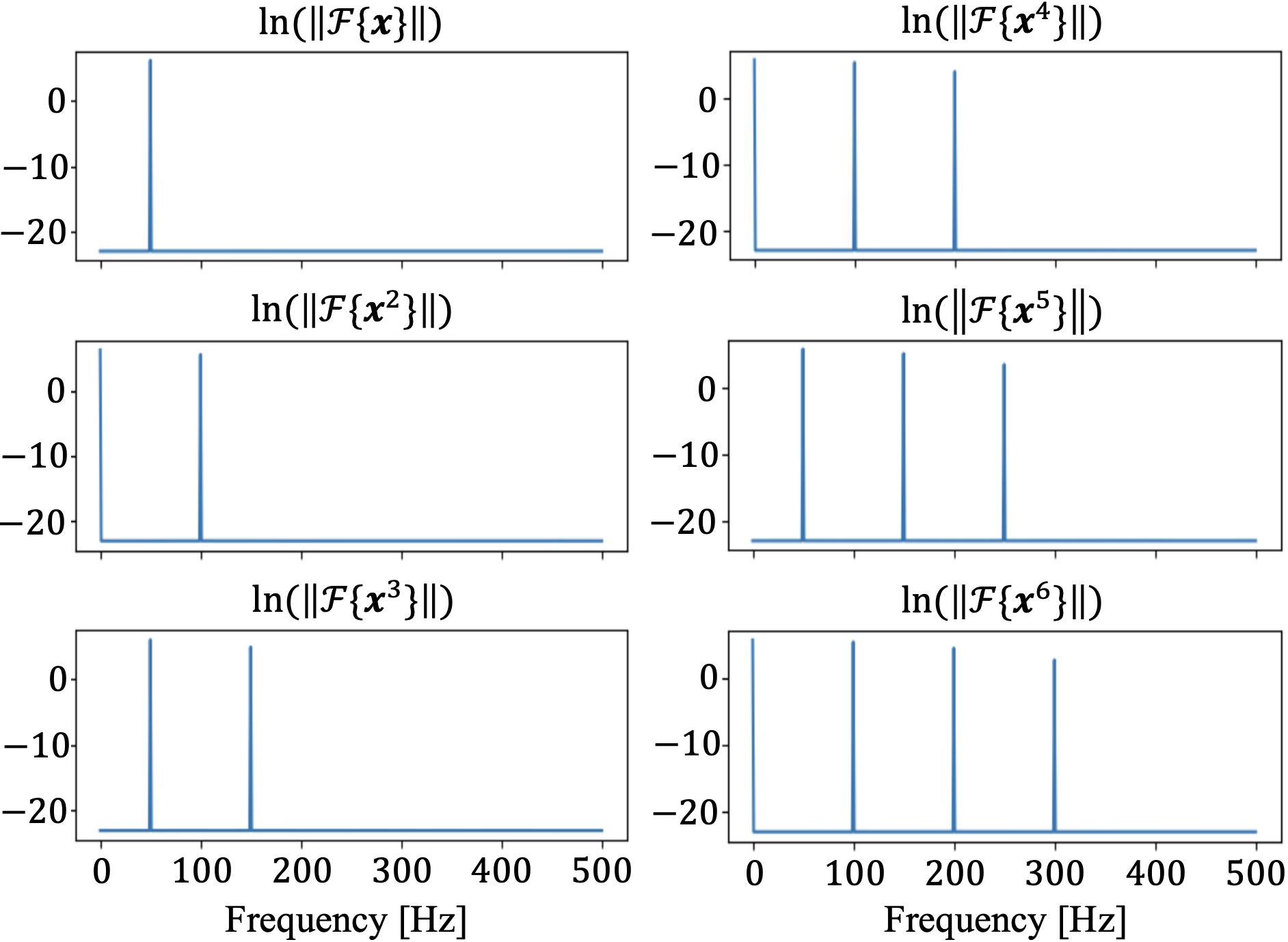}
    \caption{\small Log amplitude spectra of $\bm{x}^k$ for $k$ up to 6, where $\bm{x}$ is a 60 Hz sinusoidal signal, are shown. The sampling frequency is 1 kHz, with the discrete Fourier transform over a duration of 1 second.}
    \label{fig: powers}
    \end{center}
    \vspace{-2mm}
\end{figure}

Our hypothesis is that time-domain neural vocoders can effectively generate harmonic components from \edit{periodic priors} by modulating them through nonlinear operations.
To substantiate this hypothesis, we first examine Maclaurin expansion of a general pointwise nonlinear function, $f: \mathbb{R} \rightarrow \mathbb{R}$, assuming this function satisfies the condition for Maclaurin expansion.
Applying the Maclaurin series of $f$ to a time-domain signal $\bm{x}$ defined at each time step $n$, we obtain the following expression for the transformed signal $f(\bm{x})[n] = f(\bm{x}[n])$:
\begin{equation} \label{eq: signal Maclaurin expansion}
    f(\bm{x})[n] = \sum_{k=0}^{\infty} \, \frac{f^{(k)}(0)}{k!} \, \bm{x}^k[n].
\end{equation}
This expansion shows that $f(\bm{x})$ is a weighted sum of powers of $\bm{x}^k$, and the magnitude of the derivatives $f^{(k)}$ influences the degree of aliasing.
According to the inverse convolution theorem, the powers of a signal in the time domain correspond to repeated convolutions in the frequency domain, implying that higher-order terms introduce additional frequency components.
Figure~\ref{fig: powers} illustrates the frequency characteristics of each power up to the 6th order when $\bm{x}$ is a sinusoidal signal.

In particular, when we consider a simple periodic input signal, such as a sinusoidal signal defined as $\bm{x}[n] = \mathrm{sin}(\omega n)$, where $\omega$ is the angular frequency, the harmonics generated by the nonlinear operation can be analytically derived as shown in the following general formula:
\begin{equation} \label{eq: sine k-th power}
  \mathrm{sin}^k(\omega n) = \sum_{m=0}^k \bigl\lbrace a_m \mathrm{sin}((m\omega) n) + b_m \mathrm{cos}((m\omega) n) \bigl\rbrace,
\end{equation}
where $a_m$ and $b_m$ are coefficients dependent on $k$.
This indicates that if the input $\bm{x}$ is a sinusoidal signal, the $k$-th power term in Eq.~(\ref{eq: signal Maclaurin expansion}) becomes a sum of harmonics up to the $k$-th order, implying that nonlinear operations effectively generate harmonics that align with the input periodic signal.
\edit{The proof of Eq.~(\ref{eq: sine k-th power}) is provided in Appendix~\ref{apd: proof}.
This demonstrates why periodic priors (even for simple sinusoidal signals) offer a strong inductive bias for robust and efficient speech waveform generation when combined with time-domain nonlinear operations.}

\subsubsection{Periodic prior for time-frequency domain models}
\label{sssec: prior for tf-domain models}

Wavehax operates in the time-frequency domain, which presents an additional challenge: the effective harmonic generation mechanism through time-domain nonlinear operations is inherently unavailable.
Given the structural differences compared to time-domain vocoders, we argue that directly feeding harmonic information into the neural network is crucial for high-fidelity waveform synthesis.
\edit{To this end, we utilize a complex spectrogram derived from a harmonic prior}, denoted as $\bm{e}$, via STFT.
We construct $\bm{e}$ based on $\text{F}_0$ values $f[n]$ on each time step $n$, ensuring that the process is aliasing-free while maintaining a pseudo-constant power across all time frames.
This process is expressed by the following formula:
\begin{align}
    \bm{e}[n]
    &= \sum_{k=1}^{K_n} \sqrt{\frac{0.02}{K_n}} \, \mathrm{sin}(2 \pi k \phi[n] + k \varphi) + 0.01\bm{z}[n]. \label{eq: harmonic final}
\end{align}
Here, $F_{\text{N}}$ denotes the Nyquist frequency and $\bm{z}[n]$ is a random variable sampled from a normal distribution $\mathcal{N}(\bm{0}, I)$.
The number of harmonics, $K_n$, is defined as $K_n = \lfloor F_{\text{N}} / f[n] \rfloor$, while the cumulative phase, $\phi[n]$, is given by $\phi[n] = \sum_{m=0}^{n} f[m] / F_{\text{N}}$.
The details of this process are provided in Appendix~\ref{apd: aliasing-free harmonics generation}.
This method enables smooth phase modeling of harmonic components without relying on the structural advantages typical of time-domain vocoders.
In Section~\ref{sec: exA}, we demonstrate that using harmonic priors yields significantly better results than sinusoidal priors.

\subsection{Model architecture}
\label{ssec: model architecture}

Complex spectrograms have strong correlations between harmonic components, which are broadly distributed but appear intermittently.
Additionally, the frequency elements between harmonics exhibit significant randomness, further complicating the estimation process.
To address this complexity, we employ 2D CNNs, which have proven effective for complex spectrogram estimation as demonstrated in \cite{istftnet2}.
In Wavehax, the input harmonic signal is first transformed into a complex spectrogram using STFT with a Hanning window.
A 1D convolutional layer is applied to this complex spectrogram to capture the receptive field comprehensively along the frequency axis.
Concurrently, the input acoustic feature (e.g., mel-spectrogram) is converted into a single-channel 2D map via another 1D convolutional layer, matching the dimensions of the complex spectrogram.
These features are concatenated along the channel axis to form a five-channel 2D map, which is then processed through linear and layer normalization layers \cite{layer_norm} to align with the channels in the subsequent ConvNeXt \cite{convnext} blocks.
Each ConvNeXt block comprises a depthwise CNN, layer normalization, and a GELU \cite{gelu} activation function, sandwiched between two pointwise CNNs, which modulate the input feature map.
After passing through these blocks, the latent features are further transformed by additional layer normalization and linear layers, yielding a two-channel 2D map representing the real and imaginary components of the complex spectrogram.
Finally, the output audio waveform is reconstructed using iSTFT, followed by overlap-add with a Hanning window function.

\subsection{Adversarial training}
\label{ssec: adversarial training}

Wavehax employs the standard framework of GAN-based vocoders, incorporating mel-spectrogram loss, adversarial loss, and feature matching loss.
The mel-spectrogram loss is formulated using a function $\mathcal{M}$ that converts the input waveform into the corresponding mel-spectrogram:
\begin{equation}
    \mathcal{L}_{\text{mel}} = \| \mathcal{M}\{ \bm{x} \} - \mathcal{M} \{\mathcal{G} \{ \bm{c} \} \} \|_{1},
\end{equation}
where $\bm{x}$ denotes the natural sample from the data distribution, and $\bm{c}$ represents to acoustic features associated with $\bm{x}$, used for conditioning the generator network $\mathcal{G}$.
We adopt the hinge GAN objective \cite{soundstream}, which has shown better performance for our method than the least squares GAN objective \cite{lsgan}.
The discriminator $\mathcal{D}$ comprises multiple subdiscriminators $\mathcal{D}_k$ for $k = 1, \ldots, K$.
Each subdiscriminator identifies natural samples as real and generated samples as fake.
The optimization criterion for the discriminator is:
\begin{equation} \label{eq: D}
\begin{split}
\mathcal{L}_{\mathcal{D}} = \sum_{k=1}^{K} \max \{ \bm{0}, \bm{1} - \mathcal{D}_k \{ \bm{x} \} \} + \max \{ \bm{0}, \bm{1} + \mathcal{D}_k \{ \mathcal{G} \{ \bm{c} \} \}.
\end{split}
\end{equation}
The generator $\mathcal{G}$ aims to deceive the discriminator by minimizing the adversarial loss:
\begin{equation}
\mathcal{L}_{\text{adv}} = \sum_{k=1}^{K} \max \{ \bm{0}, \bm{1} -\mathcal{D}_k \{ \mathcal{G} \{ \bm{c} \} \} \}.
\end{equation}
The feature matching loss is the sum of the L1 distances between output latent features of each layer of each subdiscriminator.
Denoting the feature map of the $l$-th layer of the $k$-th subdiscriminator as $\mathcal{D}_{k}^l(\cdot)$, the loss is formulated as:
\begin{equation}
    \mathcal{L}_{\text{fm}} = \sum_{k}^{K} \sum_{l}^{L_k} \| \mathcal{D}_{k}^l \{ \bm{x} \} - \mathcal{D}_{k}^l \{ \mathcal{G} \{ \bm{c} \} \} \|_{1},
\end{equation}
where $L_k$ denote the number of layers in $\mathcal{D}_{k}$.
The final generator loss is defined as the sum of $\mathcal{L}_{\text{mel}}$, $\mathcal{L}_{\text{adv}}$, and $\mathcal{L}_{\text{fm}}$:
\begin{equation}
\mathcal{L}_{\mathcal{G}} = \lambda_{\text{mel}}\mathcal{L}_{\text{mel}} + \lambda_{\text{adv}} \mathcal{L}_{\text{adv}} + \lambda_{\text{fm}} \mathcal{L}_{\text{fm}},
\end{equation}
where $\lambda_{\text{mel}}$, $\lambda_{\text{adv}}$, and $\lambda_{\text{fm}}$ are balancing weights set to 45.0, 1.0, and 2.0, respectively, based on \cite{hifigan}.
To maximize the performance, we adopt the combination of HiFi-GAN's multi-period discriminator and UnivNet's multi-resolution discriminator, as suggested in \cite{univnet, bigvgan}.


\section{Experimental Evaluation}
\label{sec: exA}

We evaluate several neural vocoders through speech analysis and synthesis.
As discussed in Section~\ref{sec: background}, aliasing impacts are theoretically more pronounced when generating speech with unknown $\text{F}_0$, particularly at higher $\text{F}_0$ values.
To examine this hypothesis, we investigate performance using training data constrained to a limited $\text{F}_0$ range, allowing for explicit evaluation on unknown $\text{F}_0$ values.
We begin by presenting an ablation study on our proposed method, followed by a comparison with baseline models based on existing approaches.

\subsection{Data preparation}
\label{ssec: exA data preparation}

We used the Japanese Versatile Speech (JVS) corpus \cite{jvs}, which consists of approximately 15K utterances from 100 male and female speakers, with a total duration of about 30.2 hours.
The corpus includes normal, whispered, and falsetto speech recordings, all sampled at 24 kHz.
We used the raw audio data without volume normalization or preprocessing.
First, we carefully examined the $\text{F}_0$ range for each speaking style and speaker.
The overall $\text{F}_0$ range in the corpus was 57 to 873 Hz.
We calculated the lower 10\% and upper 20\% boundaries on a log $\text{F}_0$ scale, corresponding to frequencies below 71 Hz and above 499 Hz.
We excluded utterances from the training dataset that contained any $\text{F}_0$ outside the training range 71 $\sim$ 499 Hz.
Utterances with $\text{F}_0$ values outside the 71–499 Hz range were excluded from the training dataset, resulting in the removal of 26\% of the data for values below the lower limit and 5\% for values above the upper limit.
The evaluation dataset consisted of the following three subsets:
\begin{enumerate}[label=\Roman*.]
    \item 200 utterances with $\text{F}_0$ entirely within the training range.
    \item 200 utterances containing $\text{F}_0$ values below 71 Hz.
    \item 200 utterances containing $\text{F}_0$ values above 499 Hz.
\end{enumerate}
200 utterances, not used in the training or evaluation sets, were randomly selected for the validation dataset.
For conditioning the vocoders, we extracted 100-band log mel-spectrograms using a 2048-point Fast Fourier Transform (FFT), a Hanning window, and a 10 ms frameshift.
The mel-filter bank covered a frequency range from 0 to 8 kHz.
Additionally, we extracted $\text{F}_0$ using the Harvest algorithm \cite{harvest}, with search ranges tailored for each speaker and speaking style.

\subsection{Model details}
\label{ssec: exA model details}

We used three baseline methods, each of which was optionally extended with specific features as follows:
\begin{itemize}[left=0pt]

\item PWG \cite{pwg}: This model operates at a fixed time resolution and does not involve upsampling, which allows us to evaluate aliasing effects that are caused exclusively by the nonlinear functions.
Additionally, we explored the effectiveness of the \edit{periodic priors} and anti-aliased nonlinear operation \cite{bigvgan} in scenarios involving $\text{F}_0$ extrapolation.
We used the official implementation of the anti-aliased nonlinear operation\footnote{BigVGAN official code: \url{https://github.com/NVIDIA/BigVGAN}} for all activation layers.

\item HiFi-GAN \cite{hifigan}: This model utilizes progressive upsampling, which introduces aliasing from both upsampling and nonlinear operations.
Similar to PWG, we investigated the effectiveness of the \edit{periodic priors} and anti-aliased nonlinear operation.
Periodic waveforms were incorporated using downsampling CNNs, as described in \cite{harmonicnet}.

\item Vocos \cite{vocos}: This model estimates complex spectrograms without upsampling and generates waveforms using iSTFT.
The FFT size and window length were set to 960, with a Hanning window function.
\edit{Unlike Wavehax, Vocos does not incorporate any priors; instead, it uses 1D CNNs to estimate the log amplitude and implicit phase-wrapping \cite{vocos}.
To investigate whether periodic priors are also effective for Vocos, we optionally modified Vocos to convert the priors as log-amplitude and absolute-phase spectrograms, with phase values constrained to the range $( -\pi, \pi ]$.}
The backbone network then receives the sum of these latent features and the projected mel-spectrogram.

\item Wavehax: The FFT size and window length were set to 480, with a 50\% overlap between frames, compared to Vocos's 75\% overlap.
Since Wavehax employs 2D CNNs, computational complexity increases with FFT size.
Therefore, we selected a smaller FFT size to minimize complexity while maintaining sufficient frequency resolution.

\end{itemize}

All models were trained for 1M steps with a batch size of 16 and a sequence length of 7,680 samples (320 ms).
Gradient clipping with a threshold of 10.0 was applied to all models to stabilize training and accelerate convergence.
PWG models were trained using the Adam optimizer \cite{adam} with beta parameters set to $[0.5, 0.9]$.
The initial learning rate was $2.0 \times 10^{-4}$, halved every 200K steps.
PWG models were trained with the same loss functions as HiFi-GAN
to enhance performance \cite{nar-s2s-vc}.
HiFi-GAN was trained with the AdamW optimizer \cite{adamw}, using the original beta parameters of $[0.8, 0.99]$.
We found that the original learning rate decay of $0.999$ was too small with gradient clipping.
Therefore, we adopted the learning rate settings from BigVGAN \cite{bigvgan}, using an initial rate of $1.0 \times 10^{-4}$ and a decay factor of $0.9999996$.
Vocos and Wavehax were trained using the AdamW optimizer with a cosine learning rate scheduler, following the official Vocos code\footnote{Vocos official code: \url{https://github.com/charactr-platform/vocos}}.
The initial learning rate was $2.0 \times 10^{-4}$, with beta parameters set to $[0.8, 0.9]$.
We followed Vocos' hyperparameter settings for the discriminators except for the tuned loss weights for each subdiscriminator.

\begin{table*}
\centering
\vspace{-1mm}
\caption{\small This table shows the speech reconstruction performances from mel-spectrograms for the ablation study. The 'Prior' column represents the type of input waveform generated using $\text{F}_0$, while the 'Subset' column denotes the evaluation dataset detailed in Section \ref{ssec: exA data preparation}. IPW denotes implicit phase-wrapping \cite{vocos}. The best scores are highlighted in bold. H-RI-RI is the proposed method.}
\label{table: exA ablation study}
\renewcommand{\arraystretch}{1.26}
\begin{tabularx}{2\columnwidth}{p{1.1cm}|>{\centering\arraybackslash}p{1.2cm}|>{\centering\arraybackslash}p{1.8cm}|>{\centering\arraybackslash}p{1.8cm}|Y|YYYYY} \cline{1-10}
& Prior & Input type & Output type & Subset & VUV$\downarrow$ & RMSE$\downarrow$ & STFT$\downarrow$ & PESQ$\uparrow$ & UTMOS$\uparrow$ \\ \cline{1-10}
\multirow{3}{*}{N-RI-RI} & \multirow{3}{*}{Noise}
 & \multirow{3}{*}{\begin{tabular}{@{}c@{}}Real\\ Imaginary\end{tabular}} & \multirow{3}{*}{\begin{tabular}{@{}c@{}}Real\\ Imaginary\end{tabular}}
                 & I   & 10            & 0.111            & 0.711            & 3.108            & 2.038 \\
 & & &           & II  & 34            & 0.212            & 0.821            & 2.288            & 1.465 \\
 & & &           & III & $\textbf{4}$  & 0.078            & 0.763            & 3.306            & 1.424 \\ \cline{1-10}
\multirow{3}{*}{S-RI-RI} & \multirow{3}{*}{Sine}
 & \multirow{3}{*}{\begin{tabular}{@{}c@{}}Real\\ Imaginary\end{tabular}} & \multirow{3}{*}{\begin{tabular}{@{}c@{}}Real\\ Imaginary\end{tabular}}
                 & I   & $\textbf{6}$  & 0.085            & 0.701            & 3.733            & 2.972 \\
 & & &           & II  & $\textbf{13}$ & 0.127            & 0.708            & 3.589            & 3.326 \\
 & & &           & III & $\textbf{4}$  & 0.072            & 0.839            & 3.209            & 1.524 \\ \cline{1-10}
\multirow{3}{*}{\underline{H-RI-RI}} & \multirow{3}{*}{Harmonic}
 & \multirow{3}{*}{\begin{tabular}{@{}c@{}}Real\\ Imaginary\end{tabular}} & \multirow{3}{*}{\begin{tabular}{@{}c@{}}Real\\ Imaginary\end{tabular}}
                 & I   & $\textbf{6}$  & 0.081            & $\textbf{0.678}$ & $\textbf{3.818}$ & $\textbf{3.122}$ \\
 & & &           & II  & $\textbf{13}$ & $\textbf{0.114}$ & $\textbf{0.683}$ & $\textbf{3.702}$ & $\textbf{3.585}$ \\
 & & &           & III & $\textbf{4}$  & 0.071            & 0.722            & 3.927
      & $\textbf{1.686}$ \\ \cline{1-10}
\multirow{3}{*}{H-LA-LP} & \multirow{3}{*}{Harmonic} & \multirow{3}{*}{\begin{tabular}{@{}c@{}}Log amplitude\\ Absolute phase\end{tabular}} & \multirow{3}{*}{\begin{tabular}{@{}c@{}}Log amplitude\\ IPW\end{tabular}}
                 & I   & $\textbf{6}$  & $\textbf{0.080}$ & 0.682            & 3.793            & 3.071 \\
 & & &           & II  & $\textbf{13}$ & 0.117            & 0.687            & 3.684            & 3.531 \\
 & & &           & III & $\textbf{4}$  & $\textbf{0.070}$ & $\textbf{0.714}$ & $\textbf{3.934}$ & $\textbf{1.686}$ \\ \cline{1-10}
\multirow{3}{*}{H-L-LP} & \multirow{3}{*}{Harmonic} & \multirow{3}{*}{Log amplitude} & \multirow{3}{*}{\begin{tabular}{@{}c@{}}Log amplitude\\ IPW\end{tabular}}
                 & I   & 8             & 0.098            & 0.683            & 3.47            & 2.309 \\
 & & &           & II  & 22            & 0.167            & 0.712            & 2.84            & 1.994 \\
 & & &           & III & 4             & 0.075            & 0.720            & 3.67            & 1.524 \\ \cline{1-10}
\end{tabularx}
\end{table*}

\subsection{Evaluation metrics}
\label{ssec: evaluation metricsy}

We used six evaluation metrics, detailed below:
\begin{itemize}[left=0pt]

    \item VUV$\downarrow$ measures the percentage of incorrect classifications of speech segments as voiced or unvoiced (i.e., produced with or without vocal fold vibration), based on $\text{F}_0$ sequences extracted from recorded and synthesized speech.

    \item RMSE$\downarrow$ measures the root mean squared error of log $\text{F}_0$ values between recorded and synthesized speech, assessing $\text{F}_0$ reproduction accuracy.

    \item STFT$\downarrow$ measures the L1 distance between multi-resolution log amplitude spectrograms of recorded and synthesized speech \cite{pwg}, capturing fine and coarse spectral details.
    It is particularly useful for evaluating aliasing artifacts, which often appear as spectral distortions or blurring.
    We used FFT sizes of 512, 1024, and 2048, with corresponding window lengths and one-fourth frame shifts.

    \item PESQ$\uparrow$ assesses speech quality by comparing the spectral distance between recorded and synthesized speech, correlating with human auditory perception. We used the open-source Python library\footnote{PyPESQ: \url{https://github.com/vBaiCai/python-pesq}} for calculations.

    \item UTMOS$\uparrow$ indicates estimated subjective speech quality by a DNN model trained to predict mean opinion scores (MOS) \cite{utmos}.
    Calculations were performed using the official library\footnote{UTMOS: \url{https://github.com/sarulab-speech/UTMOS22}}.

    \item MOS$\uparrow$ measures subjective speech quality through human listener evaluations.
    Participants rated the overall naturalness of the speech on a scale from 1 (poor) to 5 (excellent), considering sound quality, clarity, and intelligibility.

\end{itemize}
All scores were calculated after normalizing each audio sample to -24 dB using the open-source Python library\footnote{pyloudnorm: \url{https://github.com/csteinmetz1/pyloudnorm}}.
PESQ and UTMOS were computed at a 16 kHz sampling rate, excluding the frequency band above 8 kHz, where aliasing is more prominent.
Subjective evaluations were conducted independently for each evaluation subset, with 30 Japanese participants recruited for each experiment, yielding a total of 300 samples per entry.

\subsection{Ablation study}
\label{ssec: exA ablation study}

Before comparing with baseline models, we present an ablation study on the proposed method.
This study examines the impact of variations in prior spectrogram types and the representation of complex spectrograms, assessed using objective metrics.
All models were trained as described in Section~\ref{ssec: exA model details}, and are outlined as follows:
\begin{itemize}[left=0pt]
    \item N-RI-RI: This model uses a \edit{non-informative prior (i.e., Gaussian noise)}.
    Both the input and output spectrograms are represented by their real and imaginary components.

    \item S-RI-RI: This model employs \edit{a sinusoidal prior}, containing only $\text{F}_0$ components.
    The input and output spectrograms are also represented by their real and imaginary components.

    \item H-RI-RI: This model uses \edit{the harmonic prior}, as detailed in Appendix~\ref{apd: aliasing-free harmonics generation}.
    The input and output spectrograms are represented by their real and imaginary components.

    \item H-LA-LP: This model also uses \edit{the harmonic prior}.
    The input spectrograms are represented in the log amplitude and absolute phase format.
    The output spectrograms are provided as pairs of log amplitude and implicit phase-wrapping, as in Vocos.

    \item H-L-LP: Unlike H-LA-LP, this model uses only the log amplitude spectrogram of the harmonic prior.

\end{itemize}

Table~\ref{table: exA ablation study} shows the evaluation results.
First, N-RI-RI and S-RI-RI exhibit significantly poorer scores compared to H-RI-RI.
This supports the hypothesis in Section \ref{ssec: harmonic prior} that time-frequency-domain processing lacks the inductive bias needed for harmonic generation, unlike time-domain processing, and that the harmonic prior significantly improves performance.
Additionally, H-L-LP, which lacks the phase information of input harmonic signals, performs worse than H-RI-RI and H-LA-LP, both of which incorporate phase information in their priors.
This suggests that time-frequency processing struggles to model smooth and coherent harmonic phases due to a lack of structural support.
H-RI-RI outperformed H-LA-LP and we adopted H-RI-RI as our final proposed model.
This approach contrasts with previous vocoders based on complex spectrogram estimation \cite{hinet, istftnet2, hiftnet, apnet, vocos}, which aim to estimate amplitude and phase pairs.
We speculate that the inherent uncertainty of $2\pi$ phase rotation complicates phase spectrogram estimation in our method.

\begin{table}
\caption{\small This table summarizes the number of learnable parameters, giga-unit multiply-accumulate operations (MACs) per second of waveform generation, and real-time factors (RTFs) measured on a single GPU (GeForce RTX 3090) and CPU (AMD EPYC 7542) with four threads. Models marked with an asterisk (*) incorporate the anti-aliased nonlinear operation \cite{bigvgan} in all activation functions. 'Har.' refers to the harmonic prior described in Appendix~\ref{apd: aliasing-free harmonics generation}. RTFs are averaged over 200 utterances.}
\label{table: model efficiency}
\renewcommand{\arraystretch}{1.26}
\begin{tabularx}{1.0\columnwidth}{p{1.4cm}>{\centering\arraybackslash}p{0.9cm}YYYY} \cline{1-6}
 & Prior & Params$\downarrow$ & MACs$\downarrow$ & GPU$\downarrow$ & CPU$\downarrow$ \\ \cline{1-6}
PWG       & Noise & 1.423 M & 33.41 & 0.0076 & 2.598 \\
PWG*      & Noise & 1.423 M & 36.53 & 0.0089 & 5.884 \\
HiFi-GAN  & --    & 13.82 M & 28.01 & 0.0051 & 0.825 \\
HiFi-GAN* & --    & 13.82 M & 29.69 & 0.0074 & 1.967 \\
Vocos     & --    & 13.49 M & 46.91 & 0.0026 & 0.036 \\
Wavehax   & Noise & 0.623 M & 1.298 & 0.0035 & 0.150 \\
Wavehax   & Har.  & 0.623 M & 1.298 & 0.0039 & 0.196 \\ \cline{1-6}
\end{tabularx}
\end{table}

\subsection{Comparison with baselines}
\label{ssec: exA comparison with baselines}

\subsubsection{Model efficiency}
\label{sssec: exA model efficiency}

Table~\ref{table: model efficiency} shows the number of trainable parameters, multiply-accumulate operations (MACs) computed using the $\mathrm{profile_macs}$ function in the torchprofile library\footnote{torchprofile: \url{https://github.com/zhijian-liu/torchprofile}}, and waveform generation speeds, expressed as real-time factors (RTF) for both GPU and CPU.
Note that the anti-aliasing mechanism involves resampling and low-pass filtering with fixed parameters, without increasing the learnable parameters.
Wavehax (i.e., H-RI-RI) retains less than 4.5\% of the parameters of HiFi-GAN and around 4.6\% of Vocos, achieving the lowest MAC count.
While the CPU generation speed of Wavehax with the harmonic prior is about 18\% of Vocos’, it outperforms HiFi-GAN by over four times.
Comparing the Wavehax models with noise or harmonic priors, the computation time for generating the harmonic prior accounts for about 23\% of the total generation time.
This is because the superposition of sinusoidal waveforms takes large computational costs.
PWG* and HiFi-GAN*, which are equipped with anti-aliased nonlinear operations, exhibit significant reductions in CPU generation speed, highlighting their impracticality in environments without GPU resources.

\subsubsection{Reconstruction performances}
\label{sssec: exA performances}

\begin{table*}
\centering
\caption{\small This table presents the speech reconstruction performances from mel-spectrograms. The 'Prior' column indicates the type of input signal generated using $\text{F}_0$. When no prior is specified, the models use only mel-spectrograms, following their original architectures. The 'Subset' column refers to the evaluation dataset described in Section~\ref{ssec: exA data preparation}.
Models marked with an asterisk (*) are equipped with anti-aliased nonlinear operations \cite{bigvgan}.
Variation in UTMOS \cite{utmos} across subsets is influenced by different speaking styles, such as whispered speech (Subset I) and falsetto speech (Subset III), which generally receive lower UTMOS scores than normal speech.
This inherent bias in recorded speech also affects the synthesized speech.
The best scores are highlighted in bold, and scores with no statistically significant difference from the best are underlined.}
\label{table: jvs evaluation}
\renewcommand{\arraystretch}{1.3}
\begin{tabularx}{2\columnwidth}{p{1.5cm}|Y|Y|YYYYY>{\centering\arraybackslash}p{1.8cm}} \cline{1-9}
 & Prior & Subset & VUV$\downarrow$ & RMSE$\downarrow$ & STFT$\downarrow$ & PESQ$\uparrow$ & UTMOS$\uparrow$ & MOS$\uparrow$ \\ \cline{1-9}
\multirow{3}{*}{Recordings}
 & \multirow{3}{*}{--}       & I   & -- & --    & --   & --   & 3.349 & 4.007 $\pm$ 0.030 \\
 &                           & II  & -- & --    & --   & --   & 3.821 & 4.127 $\pm$ 0.028 \\
 &                           & III & -- & --    & --   & --   & 1.796 & 4.033 $\pm$ 0.034 \\ \cline{1-9} \noalign{\vspace{1.8pt}} \cline{1-9}
\multirow{9}{*}{PWG}
 & \multirow{3}{*}{Noise}    & I   & $\textbf{6}$  & 0.094            & 0.725            & 3.222            & 2.418            & 2.860 $\pm$ 0.037 \\
 &                           & II  & 16            & 0.146            & 0.731            & 3.077            & 2.635            & 2.037 $\pm$ 0.032 \\
 &                           & III & 5             & 0.082            & 0.919            & 2.503            & 1.416            & 1.907 $\pm$ 0.029 \\ \cline{2-9}
 & \multirow{3}{*}{Sine}     & I   & $\textbf{6}$  & $\textbf{0.078}$ & 0.714            & 3.631            & 3.086            & 3.803 $\pm$ 0.031 \\
 &                           & II  & $\textbf{13}$ & 0.111            & 0.714            & 3.582            & 3.533            & 3.810 $\pm$ 0.031 \\
 &                           & III & $\textbf{4}$  & $\textbf{0.066}$ & 0.777            & 3.313            & 1.586            & 3.383 $\pm$ 0.033 \\ \cline{2-9}
 & \multirow{3}{*}{Harmonic} & I   & $\textbf{6}$  & 0.085            & 0.709            & 3.656            & 3.100            & 3.843 $\pm$ 0.030 \\
 &                           & II  & 14            & 0.119            & 0.711            & 3.586            & 3.552            & 3.960 $\pm$ 0.029 \\
 &                           & III & $\textbf{4}$  & 0.072            & 0.774            & 3.364            & 1.625            & 3.420 $\pm$ 0.035 \\ \cline{1-9}
\multirow{9}{*}{PWG*}
 & \multirow{3}{*}{Noise}    & I   & 7             & 0.096            & 0.713            & 3.308            & 2.536            & 3.020 $\pm$ 0.036 \\
 &                           & II  & 16            & 0.144            & 0.720            & 3.172            & 2.763            & 2.333 $\pm$ 0.033 \\
 &                           & III & 5             & 0.082            & 0.897            & 2.642            & 1.419            & 1.920 $\pm$ 0.031 \\ \cline{2-9}
 & \multirow{3}{*}{Sine}     & I   & 7             & 0.085            & 0.713            & 3.529            & 3.010            & 3.727 $\pm$ 0.033 \\
 &                           & II  & 15            & 0.125            & 0.713            & 3.481            & 3.462            & 3.603 $\pm$ 0.030 \\
 &                           & III & $\textbf{4}$  & 0.071            & 0.843            & 2.971            & 1.524            & 2.890 $\pm$ 0.036 \\ \cline{2-9}
 & \multirow{3}{*}{Harmonic} & I   & 7             & 0.083            & 0.715            & 3.605            & 3.092            & 3.853 $\pm$ 0.030 \\
 &                           & II  & 14            & 0.121            & 0.714            & 3.549            & 3.537            & 3.657 $\pm$ 0.030 \\
 &                           & III & $\textbf{4}$  & 0.071            & 0.799            & 3.229            & 1.612            & 3.140 $\pm$ 0.035 \\ \cline{1-9}
\multirow{9}{*}{HiFi-GAN}
 & \multirow{3}{*}{--}       & I   & 7             & 0.089            & 0.678            & 3.649            & 2.875            & 3.570 $\pm$ 0.034 \\
 &                           & II  & 15            & 0.136            & 0.686            & 3.464            & 3.107            & 2.803 $\pm$ 0.036 \\
 &                           & III & $\textbf{4}$  & 0.081            & 0.804            & 3.124            & 1.482            & 2.420 $\pm$ 0.034 \\ \cline{2-9}
 & \multirow{3}{*}{Sine}     & I   & $\textbf{6}$  & 0.080            & 0.666            & 3.930            & 3.195            & \underline{3.930} $\pm$ 0.031 \\
 &                           & II  & 14            & 0.113            & 0.673            & 3.807            & 3.590            & 3.877 $\pm$ 0.031 \\
 &                           & III & $\textbf{4}$  & 0.072            & 0.749            & 3.648            & 1.660            & 3.203 $\pm$ 0.037 \\ \cline{2-9}
 & \multirow{3}{*}{Harmonic} & I   & $\textbf{6}$  & 0.079            & 0.665            & 3.934            & $\textbf{3.217}$ & \underline{3.950} $\pm$ 0.031 \\
 &                           & II  & 14            & $\textbf{0.110}$ & 0.669            & 3.832            & 3.647            & \underline{3.970} $\pm$ 0.029 \\
 &                           & III & $\textbf{4}$  & 0.073            & 0.727            & 3.709            & 1.691            & 3.343 $\pm$ 0.036 \\ \cline{1-9}
\multirow{9}{*}{HiFi-GAN*}
 & \multirow{3}{*}{--}       & I   & $\textbf{6}$  & 0.087            & 0.674            & 3.703            & 2.912            & 3.543 $\pm$ 0.032 \\
 &                           & II  & 14            & 0.126            & 0.683            & 3.538            & 3.243            & 3.160 $\pm$ 0.034 \\
 &                           & III & $\textbf{4}$  & 0.082            & 0.792            & 3.162            & 1.510            & 2.507 $\pm$ 0.034 \\ \cline{2-9}
 & \multirow{3}{*}{Sine}     & I   & 7             & 0.080            & $\textbf{0.664}$ & $\textbf{3.950}$ & 3.207            & \underline{3.923} $\pm$ 0.029 \\
 &                           & II  & 15            & 0.115            & 0.700            & $\textbf{3.857}$ & $\textbf{3.649}$ & $\textbf{4.023}$ $\pm$ 0.030 \\
 &                           & III & $\textbf{4}$  & 0.071            & 0.769            & 3.517            & 1.664            & 3.027 $\pm$ 0.038 \\ \cline{2-9}
 & \multirow{3}{*}{Harmonic} & I   & 7             & 0.081            & 0.668            & 3.937            & 3.195            & $\textbf{3.973}$ $\pm$ 0.029 \\
 &                           & II  & 14            & 0.115            & $\textbf{0.673}$ & 3.826            & 3.640            & $\textbf{4.023}$ $\pm$ 0.029 \\
 &                           & III & $\textbf{4}$  & 0.072            & 0.727            & 3.749            & $\textbf{1.710}$ & 3.430 $\pm$ 0.036 \\ \cline{1-9}
\multirow{9}{*}{Vocos}
 & \multirow{3}{*}{--}       & I   & 7             & 0.090            & 0.685            & 3.656            & 2.909            & 3.520 $\pm$ 0.033 \\
 &                           & II  & 16            & 0.137            & 0.703            & 3.463            & 3.092            & 2.733 $\pm$ 0.034 \\
 &                           & III & $\textbf{4}$  & 0.077            & 0.731            & 3.456            & 1.538            & 2.683 $\pm$ 0.035 \\ \cline{2-9}
 & \multirow{3}{*}{Sine}     & I   & 7             & 0.092            & 0.695            & 3.633            & 2.904            & 3.623 $\pm$ 0.034 \\
 &                           & II  & 16            & 0.140            & 0.709            & 3.378            & 3.049            & 2.817 $\pm$ 0.035 \\
 &                           & III & $\textbf{4}$  & 0.078            & 0.735            & 3.420            & 1.545            & 2.690 $\pm$ 0.037 \\ \cline{2-9}
 & \multirow{3}{*}{Harmonic} & I   & 7             & 0.091            & 0.696            & 3.576            & 2.936            & 3.640 $\pm$ 0.032 \\
 &                           & II  & 17            & 0.136            & 0.706            & 3.394            & 3.205            & 3.197 $\pm$ 0.034 \\
 &                           & III & $\textbf{4}$  & 0.079            & 0.762            & 3.230            & 1.522            & 2.570 $\pm$ 0.037 \\ \cline{1-9} \cline{1-9}
\multirow{3}{*}{Wavehax}
 & \multirow{3}{*}{Harmonic} & I   & $\textbf{6}$  & 0.081            & 0.678            & 3.818            & 3.122            & 3.907 $\pm$ 0.032 \\
 &                           & II  & $\textbf{13}$ & 0.114            & 0.683            & 3.702            & 3.585            & \underline{3.973} $\pm$ 0.031 \\
 &                           & III & $\textbf{4}$  & 0.071            & $\textbf{0.722}$ & $\textbf{3.927}$ & 1.686            & $\textbf{3.860}$ $\pm$ 0.035 \\ \cline{1-9}
\end{tabularx}
\end{table*}

\begin{figure}[t]
    \begin{center}
    \includegraphics[width=0.9\columnwidth]{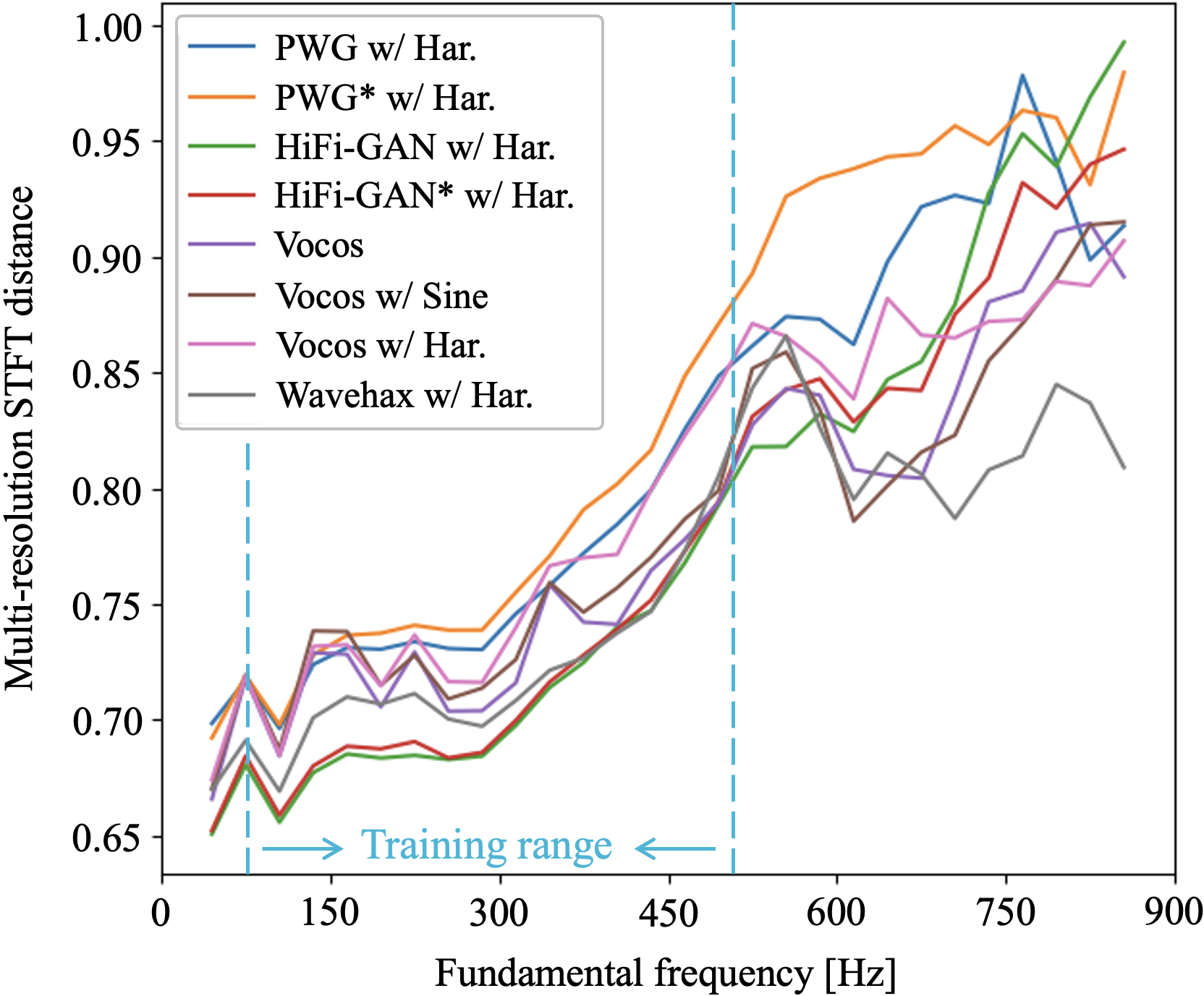}
    \end{center}
    \vspace{-1mm}
    \caption{\small This figure shows the multi-resolution STFT distances averaged for each $\text{F}_0$ bin, grouped by 30 Hz intervals.
    Models marked with an asterisk (*) are equipped with the anti-aliased nonlinear operation \cite{bigvgan}.
    'w/ Sine' and 'w/ Har.' indicate models enhanced by the sinusoidal or harmonic priors, respectively.
    The STFT frameshift is fixed at 10 ms across all resolutions to match the temporal resolution of the $\text{F}_0$ sequences.}
    \label{fig: mr-stft}
    \vspace{-2mm}
\end{figure}

Table~\ref{table: jvs evaluation} shows that Wavehax exhibits speech quality comparable to the top-performing baseline, HiFi-GAN* with harmonic prior, on subsets I and II, despite significantly fewer parameters and faster synthesis speed (see Table~\ref{table: model efficiency}).
This suggests that Wavehax avoids complicated generation processes caused by aliasing, enabling high-fidelity and efficient waveform synthesis.

On subset III, all time-domain vocoders show notable degradation, especially in MOS, with performance gaps between subsets I and III much larger than those between I and II.
This result supports the hypothesis that aliasing in neural vocoders compromises robustness in synthesizing high-pitched speech.
In contrast, Wavehax achieves significantly higher scores than all baseline models on subset III, showcasing its robustness, due to its aliasing-free design.
Figure~\ref{fig: mr-stft} further supports this, illustrating how multi-resolution STFT distances of some models change with respect to $\text{F}_0$.
The graph shows that Wavehax, unlike the other models, maintains stable performance without significant degradation at unseen high $\text{F}_0$ values.

Compared to the Vocos models, Wavehax consistently outperforms them across all metrics, particularly in subsets II and III.
This discrepancy likely arises from differences in architectural design.
Vocos relies on 1D CNNs to achieve high expressiveness and complexity, strongly reflecting the training data through a large number of parameters.
In contrast, Wavehax employs 2D CNNs, capturing time-frequency representations in more redundant latent feature spaces with fewer parameters, allowing more effective adaptation to unseen conditions.
The translational invariance of 2D CNNs further enhances robustness by preserving spatial coherence in processing harmonic spectrograms.

Focusing on the time-domain baseline models, PWG and HiFi-GAN, we observe that the harmonic prior almost consistently yields the best results.
Morrison et al. \cite{cargan} argued that non-autoregressive GAN-based vocoders struggle to model the relationship between $\text{F}_0$ and harmonic phases, which involves cumulative summation.
This challenge can also extend to harmonic components not generated via time-domain nonlinear operations, where certain terms in Eq.~(\ref{eq: signal Maclaurin expansion}) can disappear due to vanishing derivatives of particular orders.
We speculate that models using the sinusoidal prior build harmonic structures less effectively than those with harmonic priors.
Additionally, anti-aliased nonlinear operations have proven effective in some cases, such as PWG* and HiFi-GAN* without periodic priors, and HiFi-GAN* with the harmonic prior.
However, the improvements brought by them were inconsistent, implying the limitations discussed in Section~\ref{sssec: problems of anti-aliased nonlinear operation}.

\section{Conclusion}

This paper presented Wavehax, a neural vocoder designed to achieve aliasing-free audio waveform synthesis leveraging complex spectrogram estimation.
We provided a theoretical analysis of why time-frequency-domain processing inherently avoids aliasing, in contrast with time-domain processing.
Additionally, we demonstrated that combining 2D CNNs with harmonic spectrograms plays a crucial role in enabling reliable complex spectrogram estimation.
Wavehax offers substantial efficiency improvements over high-fidelity time-domain neural vocoders while maintaining comparable speech quality, by avoiding aliasing in latent spaces.
Moreover, Wavehax outperforms Vocos in terms of perceptual quality, further validating the effectiveness of our architectural design in overcoming the lack of harmonic generation bias in time-frequency-domain processing.
This work highlighted the advantages of aliasing-free waveform synthesis, paving the way for further improvements in neural vocoders.

\appendices

\section{Harmonic Decomposition of Sine Powers}
\label{apd: proof}

We prove that for any non-negative integer $n$, $\mathrm{sin}^n (\theta)$ can be expressed as a linear combination of sine and cosine terms, involving harmonics up to and including the $n$-th harmonic.
Specifically, we prove the following proposition by mathematical induction:
\begin{equation} \label{eq: proposition}
  \sin^n(\theta) = \sum_{m=0}^{n} \bigl\lbrace a_m \sin(m\theta) + b_m \cos(m\theta) \bigl\rbrace
\end{equation}
where $a_m$ and $b_m$ are coefficients dependent on $n$.

\noindent 1) For $n = 0$, the proposition holds because
\begin{equation} \label{eq: proposition n=0}
  \mathrm{sin}^0(\theta) = 1 = \mathrm{cos}(0 \, \theta).
\end{equation}
2) Assuming Eq.~(\ref{eq: proposition}) holds for $n=k$ ($k \geq 0$), we prove the case for $n=k+1$ as follows.
\begin{align} \label{eq: proposition n=k+1}
  \mathrm{sin}&^{k+1}(\theta) = \mathrm{sin}(\theta) \sum_{m=0}^k \bigl\lbrace a_m \mathrm{sin} (m\theta) + b_m \mathrm{cos}(m\theta) \bigl\rbrace \nonumber \\
  &= \sum_{m=0}^k \bigl\lbrace a_m \mathrm{sin}(m\theta) \, \mathrm{sin}(\theta) + b_m \mathrm{cos}(m\theta) \, \mathrm{sin}(\theta) \bigl\rbrace \nonumber \\
  &= \sum_{m=0}^k \Bigl\lbrace \frac{a_m}{2} \bigl( \mathrm{cos}((m-1)\theta) - \mathrm{cos}((m+1)\theta) \bigl) \nonumber \\
  &\qquad ~\, + \frac{b_m}{2} \bigl( \mathrm{sin}((m+1)\theta) - \mathrm{sin}((m-1)\theta) \bigl) \Bigl\rbrace \nonumber \\
  &= \sum_{m=0}^{k+1} \bigl\lbrace a'_m \mathrm{sin}(m\theta) + b'_m \mathrm{cos}(m\theta), \bigl\rbrace
\end{align}
where $a'_m$ and $b'_m$ are the updated coefficients derived from $a_m$ and $b_m$.
The transformation between the second and third lines is achieved using the following product-to-sum identities:
\begin{equation} \label{eq: product-to-sum identiy1}
  \mathrm{sin}(m\theta) \, \mathrm{sin}(\theta) = \frac{ \mathrm{cos}((m-1)\theta) - \mathrm{cos}((m+1)\theta) }{2}
\end{equation}
\begin{equation} \label{eq: product-to-sum identiy2}
  \mathrm{cos}(m\theta) \, \mathrm{sin}(\theta) = \frac{ \mathrm{sin}((m+1)\theta) - \mathrm{sin}((m-1)\theta) }{2}.
\end{equation}
Consequently, from steps 1) and 2), the proposition of Eq.~(\ref{eq: proposition}) has been proved.

\section{Aliasing-Free Harmonic Signal Generation}
\label{apd: aliasing-free harmonics generation}

Pulse trains are widely used as excitation signals in vocoders \cite{straight, world, nhv}, and are a natural candidate for the harmonic prior.
However, pulse trains, constructed via pointwise sampling from continuous-time signals, inherently involve aliasing.
To avoid aliasing, we adopt a superposition of band-limited sinusoidal signals as \cite{nhv}.
We extend this approach by modulating each harmonic’s amplitude to maintain constant signal power independent of $\text{F}_0$ as \cite{straight, world}.
This helps the network focus on relevant features without being affected by power variations, enhancing its generalizability to unseen $\text{F}_0$.
The harmonic signal is defined as follows:
\begin{align}
    \bm{h}[n]
    &= \sum_{k=1}^{K_n} g_k[n] \, \mathrm{sin}(2 \pi k \phi[n] + \varphi_k) \label{eq: harmonic} \\
    &= \sum_{k=-K_n}^{K_n} \frac{g_k[n]}{2} \, \mathrm{exp}(j \varphi_k) \, \mathrm{exp}(j 2 \pi k \phi[n]),
\end{align}
where $g_k[n]$ is the amplitude of the $k$-th harmonic component at time step $n$, and $\varphi_k$ is the initial phase of the $k$-th harmonic component.
$f[n]$ and $\phi[n]$ are ones defined in Section~\ref{sssec: prior for tf-domain models}.
The number of harmonic components $K_n$ is determined by dividing the maximum frequency $F_{\text{max}}$ by $f[n]$, i.e., $K_n = \lfloor F_{\text{max}} / f[n] \rfloor$.
$F_{\text{max}}$ sets the upper harmonic frequency limit, balancing frequency coverage and computational cost.
We set $F_{\text{max}}$ to the Nyquist frequency $F_{\text{N}}$.
Signal power can be controlled by modulating $g_k[n]$ based on Parseval’s identity:
\begin{equation} \label{eq: parseval identity}
    \frac{1}{L} \sum_{k=-K_n}^{K_n} \left( \frac{g_k[n]}{2} \right) ^2 = \sum_{m=n - L/2}^{n + L/2} \bm{e}[m]^2,
\end{equation}
where $L$ is the window length.
Since we assume constant power $C^2$ for each time frame, the identity becomes:
\begin{equation} \label{eq: parseval identity2}
    \frac{1}{L^2} \sum_{k=-K_n}^{K_n} \left( \frac{g_k[n]}{2} \right) ^2 = \frac{1}{L} \sum_{m=n - L/2}^{n + L/2} \bm{e}[m]^2 = C^2,
\end{equation}
While alternative spectral envelopes, such as linear decay, can be considered, this paper assumes a flat spectral envelope across each time frame (i.e., $g_k[n] = g_{k+1}[n]$ for all $k \leq K_n - 1$), yielding:
\begin{equation} \label{eq: parseval identity3}
     \sum_{k=-K_n}^{K_n} \left( \frac{g_k[n]}{2} \right) ^2 = 2K_n \left( \frac{g_k[n]}{2} \right) ^2 = L^2 C^2.
\end{equation}
\begin{equation} \label{eq: coefficients}
    g_k[n] =
    \begin{cases}
        LC \sqrt{2 / K_n} ~ & \mbox{if} ~ f[n] \neq 0 \\
        0 ~ & \mbox{otherwise}
    \end{cases}
\end{equation}
where $LC$ is set to 0.1 in this study.
Additionally, we observed better performance when using linear phases for $\varphi_k$ compared to zero or random phases.
Linear initial phases are defined as:
\begin{equation} \label{eq: initial phases}
    \varphi_k = k \varphi,
\end{equation}
where $\varphi \sim \mathcal{U}(-\pi, \pi)$.
The harmonic signal $\bm{h}$ is derived by substituting Eq.~(\ref{eq: coefficients}) and Eq.~(\ref{eq: initial phases}) into Eq.~(\ref{eq: harmonic}).

\ifCLASSOPTIONcaptionsoff
  \newpage
\fi



%



\bibliography{main}
\bibliographystyle{IEEEtran}

%








\end{document}